\documentclass[twoside,twocolumn,9pt]{article}
\usepackage{extsizes}
\usepackage[super,sort&compress,comma]{natbib} 
\usepackage[version=3]{mhchem}
\usepackage[left=1.5cm, right=1.5cm, top=1.785cm, bottom=2.0cm]{geometry}
\usepackage{balance}
\usepackage{mathptmx}
\usepackage{sectsty}
\usepackage{graphicx} 
\usepackage{lastpage}
\usepackage[format=plain,justification=justified,singlelinecheck=false,font={stretch=1.125,small,sf},labelfont=bf,labelsep=space]{caption}
\usepackage{float}
\usepackage{fancyhdr}
\usepackage{fnpos}
\usepackage[english]{babel}
\addto{\captionsenglish}{%
  
}
\usepackage{array}
\usepackage{droidsans}
\usepackage{charter}
\usepackage[T1]{fontenc}
\usepackage[usenames,dvipsnames]{xcolor}
\usepackage{setspace}
\usepackage[compact]{titlesec}
\usepackage{hyperref}

\usepackage{epstopdf}
\usepackage{textgreek} 

\definecolor{cream}{RGB}{222,217,201}

\begin{document}

\pagestyle{fancy}
\thispagestyle{plain}
\fancypagestyle{plain}{
\renewcommand{\headrulewidth}{0pt}
}

\makeFNbottom
\makeatletter
\renewcommand\LARGE{\@setfontsize\LARGE{15pt}{17}}
\renewcommand\Large{\@setfontsize\Large{12pt}{14}}
\renewcommand\large{\@setfontsize\large{10pt}{12}}
\renewcommand\footnotesize{\@setfontsize\footnotesize{7pt}{10}}
\makeatother

\renewcommand{\thefootnote}{\fnsymbol{footnote}}
\renewcommand\footnoterule{\vspace*{1pt}%
\color{cream}\hrule width 3.5in height 0.4pt \color{black}\vspace*{5pt}} 
\setcounter{secnumdepth}{5}

\makeatletter 
\renewcommand\@biblabel[1]{#1}            
\renewcommand\@makefntext[1]%
{\noindent\makebox[0pt][r]{\@thefnmark\,}#1}
\makeatother 
\renewcommand{\figurename}{\small{Fig.}~}
\sectionfont{\sffamily\Large}
\subsectionfont{\normalsize}
\subsubsectionfont{\bf}
\setstretch{1.125} 
\setlength{\skip\footins}{0.8cm}
\setlength{\footnotesep}{0.25cm}
\setlength{\jot}{10pt}
\titlespacing*{\section}{0pt}{4pt}{4pt}
\titlespacing*{\subsection}{0pt}{15pt}{1pt}



\makeatletter 
\newlength{\figrulesep} 
\setlength{\figrulesep}{0.5\textfloatsep} 

\newcommand{\topfigrule}{\vspace*{-1pt}%
\noindent{\color{cream}\rule[-\figrulesep]{\columnwidth}{1.5pt}} }

\newcommand{\botfigrule}{\vspace*{-2pt}%
\noindent{\color{cream}\rule[\figrulesep]{\columnwidth}{1.5pt}} }

\newcommand{\dblfigrule}{\vspace*{-1pt}%
\noindent{\color{cream}\rule[-\figrulesep]{\textwidth}{1.5pt}} }

\makeatother

\twocolumn[
  \begin{@twocolumnfalse}
\vspace{1em}
\sffamily

\centering
\noindent\LARGE{\textbf{Photoluminescence Quenching in $\text{WSe}_2$ via p-Doping Induced by Functionalized Rylene Dyes}} \\
\vspace{0.3cm} 

\noindent\large{Ana M. Valencia,\textit{$^{a}$} Theresa Kuechle,\textit{$^{b}$} Maximiliam Tomoscheit,\textit{$^{b}$} Sarah Jasmin Finkelmeyer,\textit{$^{c}$} Olga Utismenko,\textit{$^{d}$}, Kalina Peneva,\textit{$^{d,e}$} Martin Presselt,\textit{$^{c,e,f}$} Giancarlo Soavi,\textit{$^{b,g}$} and Caterina Cocchi,\textit{$^{a,h}$}}

\vspace{0.5cm}

\noindent\normalsize{Hybrid heterostructures combining transition metal dichalcogenides (TMDs) with light-harvesting dyes are promising materials for next-generation optoelectronics. Yet, controlling and understanding interfacial charge transfer mechanisms in these complex systems remains a major challenge. Here, we investigate the microscopic origin of photoluminescence (PL) quenching in $\text{WSe}_2$ functionalized with a novel, strongly electron-deficient perylene monoimide dye, $\text{CN}_4\text{PMI}$. Experimentally, the hybridization induces a $\sim$97\% PL quenching in $\text{WSe}_2$, confirming substantial static charge transfer and increased $p$-doping from the dye. To isolate the dominant electronic mechanism, we investigate from first principles various interface morphologies, including differing molecular orientations and layer thicknesses. Our density-functional theory results confirm that $\text{CN}_4\text{PMI}$ acts as a strong electron acceptor, inducing $p$-doping and forming a type-II level alignment with all considered configurations, giving rise to a small or vanishing band gap. Based on these findings, we attribute the observed PL suppression in $\text{WSe}_2$ to these strong electronic interactions with the dye. Our study provides a clear and validated strategy for tailoring the electronic structure of TMDs through targeted, electron-deficient organic functionalization.} \\


 \end{@twocolumnfalse} \vspace{0.6cm}]

\renewcommand*\rmdefault{bch}\normalfont\upshape
\rmfamily
\section*{}
\vspace{-1cm}


\footnotetext{\textit{$^{a}$~Institute of Physics, Carl-von-Ossietzky Universit{\"a}t Oldenburg, 26129 Oldenburg, Germany; E-mail: ana.valencia@uni-oldenburg.de}}
\footnotetext{\textit{$^{b}$~Institute of Solid State Physics, Friedrich-Schiller-Universit{\"a}t Jena, 07743 Jena, Germany.}}
\footnotetext{\textit{$^{c}$~Leibniz Institute of Photonic Technology, 07745 Jena, Germany.}}
\footnotetext{\textit{$^{d}$~Institute of Organic Chemistry and Macromolecular Chemistry, Friedrich-Schiller-Universit{\"a}t Jena, 07743 Jena, Germany.}}
\footnotetext{\textit{$^{e}$~Center for Energy and Environmental Chemistry Jena (CEEC Jena), Friedrich Schiller Universit{\"a}t Jena, 07743 Jena, Germany.}}
\footnotetext{\textit{$^{f}$~sciclus GmbH \& Co. KG, 07745 Jena, Germany.}}
\footnotetext{\textit{$^{g}$~Abbe Center of Photonics, Friedrich-Schiller-Universit{\"a}t Jena, 07743 Jena, Germany.}}
\footnotetext{\textit{$^{h}$~Institute of Condensed Matter Theory and Optics, Friedrich-Schiller-Universit{\"a}t Jena, 07743 Jena, Germany ; E-mail:caterina.cocchi@uni-jena.de.}}




\section{Introduction}
The combination of light-harvesting dyes and transition metal dichalcogenide (TMD) monolayers, an established family of low-dimensional semiconductors exhibiting excellent transport properties,\cite{eich+15acsnano,pono+18acsnano,hien+20prb} a rich exciton physics,\cite{wang+18rmp,rega+22nrm} and efficient photoluminescence (PL),\cite{sple+10nl,eda+11nl,tonn+13oe,aman+15sci,shi+162Dm} is expected to unite the best of two worlds.\cite{koch21apl,cui+24cr} These hybrid heterostructures promise a new class of materials for next-generation optoelectronic~\cite{gobb+18am,ji-choi22ns} and photonic \cite{deab+25acsp} applications. 
To achieve this ambitious goal, a deep understanding of the fundamental processes that govern the interaction between dyes and TMD is mandatory. 

Molecular adsorption is a viable tool to tune the optical properties of TMD,\cite{nguy+15am,zhen+16nano,thom+23npj2d} leading to both PL enhancement or suppression depending on the specific characteristics of the heterostructure.\cite{mour+13nl,habi+18ns,KUECHLE2021100097} On one hand, PL can be enhanced when molecular adsorbates~\cite{sim+15nano,bret+21nano,zhu+25jpcc} or solvent molecules~\cite{li+25mcf} heal detrimental defects~\cite{luha+24acsanm} within the TMD lattice. On the other hand, PL is typically suppressed, primarily due to two concurrent effects related to charge transfer: spatial separation of photoinduced charge carriers,\cite{qin+22acsph} and doping, either from a substrate~\cite{muns+24jpcl,zhu+25jpcc} or within a heterointerface,\cite{ceba+14nano,decl+23jpcc} both decreasing the probability of radiative recombination.\cite{Cadore_2024,Der-Hsien+19science} Effective charge transfer, associated with a type-II level alignment, is known to reduce the inherent TMD PL.\cite{khan+20ass,chen+23jl} Given the complexity of these competing mechanisms and the large variety of parameters that can influence the optical properties, a systematic, targeted analysis is urgently needed to isolate and understand the dominant effect.

To better understand the PL suppression scenario and investigate charge transfer mechanisms at organic-inorganic interfaces, we designed a novel perylene monoimide dye functionalized with four cyano groups (\ce{CN4PMI}) at bay positions. Cyano groups are well-known for their strong electron-withdrawing character,\cite{wang+23jmca} which significantly lowers the energy of the lowest unoccupied molecular orbital (LUMO), thus enhancing the electron affinity of the molecule.\cite{liu+02mm} This molecular design, which builds upon strategies commonly used in organic photovoltaics to tailor energy levels and enable type-II band alignment,\cite{wurt04cc,kipp-bred09ees,weil+10acie} enables \ce{CN4PMI} to efficiently accept electrons from \ce{WSe2} and promote $p$-type doping in the TMD. Furthermore, the compact and planar nature of the cyano substituents preserves $\pi$-conjugation, supporting strong $\pi-\pi$ interactions and a flat-lying adsorption geometry, both favoring interfacial charge transfer.\cite{das+18aem,fink-mart25caej}

Acknowledging the inherent experimental difficulties in synthesizing well-ordered samples and obtaining precise information about their morphology and composition, we use density-functional theory (DFT) as an ideal guide to rationalize PL quenching mechanisms measured in a hybrid heterostructure formed by \ce{WSe2} decorated by \ce{CN4PMI} molecules. We explore different configurations with varying substrate, molecular film thickness, and organic layer orientations. Molecular adsorption is stable in all considered conditions, but it is particularly favored when the dyes lie flat on the substrate. Due to its electron-acceptor nature, \ce{CN4PMI} induces strong $p$-doping on \ce{WSe2}, leading to a type-II level alignment and a narrow ($<500$ meV) or vanishing band gap. The strong electronic interactions between adsorbate and substrate, regardless of the specific characteristics of the hybrid interface, strongly point to their role as the driving mechanism for the observed PL quenching in \ce{WSe2}.

\section{Materials and Methods}

\subsection{Synthesis}
The \ce{CN4PMI} dye was synthesized via a one-pot two-step procedure starting from tetrabromo-perylene-3,4:9,10-tetracarboxylic monoanhydride (PMA-4Br). First, the precursor was subjected to nucleophilic aromatic substitution with copper(I) cyanide (CuCN) in the presence of zinc acetate in dry N-methyl-2-pyrrolidone (NMP) at 130$^{\circ}$C under an inert atmosphere. This reaction substituted the four bromine atoms at the bay positions with cyano groups, yielding a tetra-cyano-substituted perylene monoanhydride intermediate. N-octylamine was then added directly, promoting imidization under the same conditions and forming the monoimide product in situ. The crude product was isolated by precipitation, followed by purification via silica gel column chromatography. The structure of \ce{CN4PMI} was confirmed by NMR spectroscopy and high-resolution mass spectrometry, consistent with complete substitution at the bay positions and successful monoimide formation. The resulting molecule is highly planar, displays strong absorption in the visible range, and retains luminescent properties in dilute solution. Further details are provided in the Electronic Supplementary Information (ESI).

\begin{figure}
    \centering
    \includegraphics[width=0.5\textwidth]{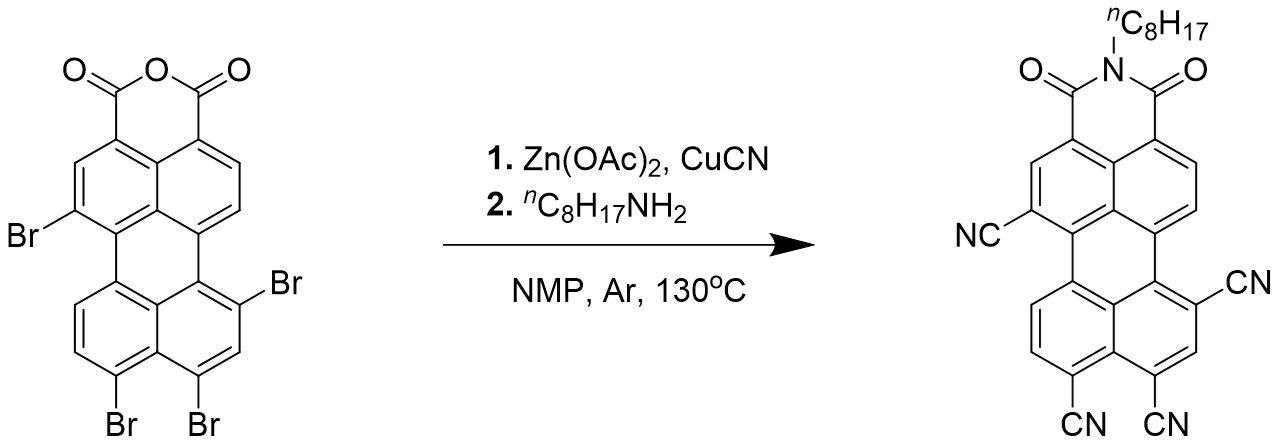}
    \caption{Synthesis of \ce{CN4PMI}.}
    \label{fig:enter-label}
\end{figure}

\subsection{Experimental Methods}
\subsubsection{Mechanical exfoliation of \ce{WSe2} monolayer}
The \ce{WSe2} flakes were fabricated by mechanical exfoliation. As a first step, bulk material was thinned down before being exfoliated one more time onto a PDMS film. Through optical microscopy contrast, areas of a single layer were identified and subsequently positioned over a Si/SiO$_2$ substrate using a mechanical xyz stage. The viscoelastic properties of the PDMS allow for a careful separation from the monolayer, while keeping it intact. To increase the likelihood of a successful transfer, the system was slowly heated to 60~$^{\circ}$C.

\subsubsection{Langmuir-Blodgett method}
The UV-Vis absorption and emission spectra of \ce{CN4PMI} in chloroform are presented in Fig.~\ref{fig:PL}a. The surface pressure [mean molecular area, $\Pi$(mma)] isotherm of \ce{CN4PMI} was recorded after spreading 1500~$\mu L$ of 0.1~$\mu$mol/mL chloroform solution kept at room temperature onto a tempered-controlled pure water subphase (25$^{\circ}$C, conductivity 0.16 to 0.25~$\mu$S/cm) in an Langmuir-Blodgett (LB) trough (KSV 5000; length = 520~mm; compression length=475.2~mm; width=150~mm). A 20-minute equilibration period was observed to ensure complete solvent evaporation before initiating barrier compression.

The barriers were moved at a maximum rate of 10~mm/min, with an additional pressure rate limit of 5~mN/m/min enforced once the surface pressure began to rise. The higher rate was applied while the surface pressure remained constant; when the surface pressure began to rise, the rate limitation was enforced. For the preparation of the hybrid heterostructures, the supported \ce{WSe2} monolayer was first immersed in pure water before spreading the dye. The spreading and compression conditions were maintained identical to those used for the isotherm measurements. Upon reaching a surface pressure of 30 mN/m, corresponding to a dense packing of the perylene molecules,\cite{hupf+21lang} the Langmuir layer was transferred onto the \ce{WSe2} monolayer via the LB technique using vertical upward withdrawal at a constant speed of 2-3~mm/min.

\subsubsection{Brewster angle microscopy (BAM)}
Brewster angle microscopy (BAM) images were recorded using a KSV NIMA MicroBAM (Biolin Scientific) before spreading (pure water subphase), after spreading, during the solvent evaporation period, and periodically throughout all stages of the \textPi(mma) isotherm measurement.\cite{finkelmeyer2025}

\subsubsection{PL setup}
For the excitation of the samples, we used a continuous wave (CW) laser system (MixTrain from Spectra-Physics) with tunable output in the wavelength range from 515 to 670~nm. The excitation beam is then coupled to a custom-built microscope via a 90:10 beam splitter, and subsequently it is focused on the sample using an objective (Mitutoyo M Plan Apo HL, 50x) which allows for obtaining a $1/e^2$ spot size at the sample position of approximately 1.59~\textmu m at 633~nm. The back-scattered PL signal is passed through the same objective and, after filtering using notch filters to remove the excitation beam, it is directed to a monochromator (Horiba iHR550) equipped with an electrically cooled CCD camera (SynapsePlus detector) for low-noise detection.

\subsection{Computational Details}
All calculations presented in this work are carried out in the framework of DFT, using the all-electron code FHI-aims~\cite{blum+09cpc} in the scalar relativistic approximation. The heterostructures were relaxed using the Perdew-Burke-Ernzerhof (PBE) exchange-correlation functional~\cite{perd+96prl} supplemented by the pairwise Tkatchenko-Scheffler scheme~\cite{TS09prl} for dispersion corrections. A 60~\AA{} thick vacuum layer and dipole correction were included to prevent spurious electrostatic interactions between periodic replicas. The electronic properties were computed with the range-separated hybrid functional HSE06~\cite{hse06erratum} including spin-orbit coupling. ``Tight'' numerical settings were adopted for the PBE calculations, while the ``LVL-intermediate'' settings were employed with HSE06. Brillouin zone integrations were performed using a Monkhorst-Pack \textbf{k}-mesh with 12$\times$12$\times$1 points for the primitive cell of \ce{WSe2} and 6$\times$3$\times$1 for the hybrid interfaces simulated in supercells. Convergence criteria were set to $10^{-6}$ eV for the total energy and $10^{-5}$ for changes in the density. Structural relaxations were carried out with a threshold of 0.001~eV/\AA{} for interatomic forces.

\section{Results}

\subsection{PL spectroscopy}
\begin{figure*}
    \centering
\includegraphics[width=\textwidth]{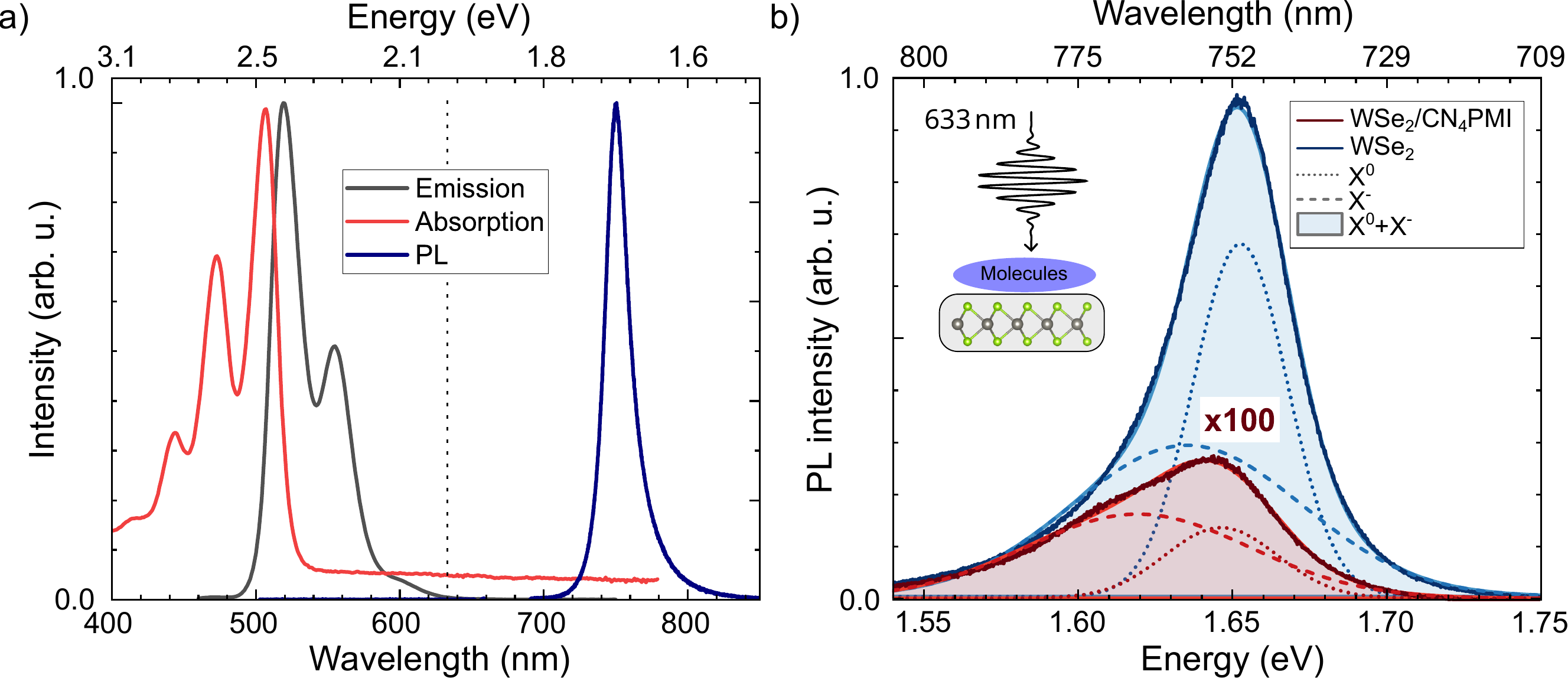}
    \caption{a) UV-Vis absorption (red curve) and emission spectrum (black curve) of \ce{CN4PMI}, both taken at room temperature. b) PL spectra of a WSe$_2$ monolayer before (blue curve) and after hybridization with \ce{CN4PMI} via LB layer coating (red curve), both excited in ambient conditions with the same laser power density of 5000 W/cm$^2$ and wavelength of 633~nm taken. X$^{0}$ and X$^{-}$ denote the neutral exciton and the trion resonances}
    \label{fig:PL}
\end{figure*}

We start our analysis by comparing the PL spectra of the hybrid \ce{CN4PMI}-\ce{WSe2} interfaces with those of a pristine \ce{WSe2} monolayer (Fig.~\ref{fig:PL}b). To ensure consistency, measurements were performed on the same \ce{WSe2} sample before and after molecular deposition. Its monolayer nature is confirmed by the identification of the typical direct gap PL signature at 1.67~eV and by the absence of any lower energy PL emission (Fig.~S8), which is instead typical of multi-layer samples.\cite{yan+14apl} The hybrid heterostructure was prepared by depositing a densely packed LB layer of \ce{CN4PMI} onto the supported TMD via vertical upward transfer. In this process, the \ce{WSe2}  monolayer, initially immersed in the pure water subphase, was withdrawn through the air-water interface after \ce{CN4PMI} dyes were spread and laterally compressed into a dense, non-covalently bonded film. This preparation method ensures a well-defined interface dominated by physisorption and strong $\pi$-$\pi$ interactions between the planar dyes and the \ce{WSe2} lattice.  

For the PL measurements, we used an excitation wavelength of 633~nm at an average power density of 5000~W/cm$^2$. The hybridization process in the heterostructure results in two main spectral changes in \ce{WSe2}. First, we observe a strong reduction of the PL intensity by $\sim97\%$, which can have both a static or dynamic origin, similar to previous findings for TMD-graphene heterostructures.\cite{Ghaebi2025} Static charge transfer due to hybridization leads to doping in the TMD, and thus to a higher trion density, which typically displays very low PL yield due to ultrafast non-radiative recombination pathways.\cite{Lien2019, Li2022} On the other hand, ultrafast charge transfer following photoexcitation can reduce the number of available electron-hole pairs, which are necessary for radiative recombination. 
To clarify the quenching mechanism and rule out alternative pathways, we compare the spectral overlap between the emission of the \ce{WSe2} monolayer and the absorption of the \ce{CN4PMI} dye in Fig.~\ref{fig:PL}a. The \ce{WSe2} PL peaks at approximately 750 nm (1.65 eV), whereas the dye absorption is confined to the 400–550 nm range. The absence of any spectral overlap between the TMD emission and the dye absorption definitively rules out long-range F\"orster resonance energy transfer as the origin of the PL quenching.

Another drastic difference in the PL spectra of the heterostucture compared to pristine \ce{WSe2} is the change in the trion density. To estimate the exciton/trion ratio, we applied a cumulative Gaussian fit to the neutral A exciton peak at higher energy (1.665~eV for the pristine sample and 1.650~eV for the hybridized sample) and the trion peak at lower energy (1.647~eV for the pristine and 1.640~eV for the hybridized sample).\cite{Wierzbowski2017} To quantify the excitonic changes, the PL spectra were deconvolved using a cumulative fit of two Gaussian functions, with the peak energies, integrated areas, and full-widths at half-maximum treated as unconstrained parameters alongside a constant baseline offset. The exciton/trion ratio, defined by the integrated PL intensities of the respective peaks, decreases from 0.88 for the pristine \ce{WSe2} to 0.28 for the hybrid \ce{CN4PMI}-\ce{WSe2} sample.

This significant increase in the trion density provides direct experimental evidence of static charge transfer from the \ce{WSe2} monolayer to the molecular adsorbates,\cite{Kong2022} in agreement with previous reports on \ce{WSe2}-perylene hybrids.\cite{Voelzer2023}. Such $p$-doping is a well-documented mechanism for PL quenching in TMDs, as the excess carriers open efficient non-radiative recombination channels. 
Furthermore, we observe a redshift by approximately 6-9~meV of the PL peaks from pristine to hybrid samples for the neutral exciton and trion, respectively, likely indicating the concomitant effect of charge transfer and dielectric screening induced by the molecular overlayer.\cite{zhu+18sa,Mak2013,Muth2024,Aly2025}

It is worth noting that the observed $\sim$97\% quenching of the PL intensity is remarkably high compared to traditional electrostatic gating configurations. For instance, studies on gated \ce{MoS2} typically show a quenching factor of only approximately 4 between opposite doping extremes,\cite{liu+22na} while \ce{WSe2} $p$-$n$ junctions often require cryogenic temperatures to achieve significant modulation.\cite{ross+14natn,shin+24am} The fact that our molecularly functionalized \ce{WSe2} sample exhibits such a large suppression at room temperature confirms that such quenching arises from a synergistic effect between static $p$-doping, evidenced by the modified trion ratio, which enhances non-radiative recombination, and electronic hybridization, which facilitates ultrafast charge transfer. The latter dynamic pathway allows the molecules to act as a physical sink for photoexcited electrons, leading to a much stronger PL suppression than that achievable through purely electrostatic means.

Overall, our experimental analysis of PL spectra in pristine \ce{WSe2} and hybrid \ce{CN4PMI}-\ce{WSe2} samples delivers two main findings: a reduction of the exciton/trion ratio and a concomitant overall quenching of the PL intensity. Both observations can be rationalized in the context of a type-II band alignment, which could be responsible for both a static charge transfer immediately after hybridization and a dynamical photo-induced doping, leading to an overall reduction of the PL intensity due to the presence of additional non-radiative recombination channels, as previously discussed in the literature of molecularly decorated TMDs.\cite{der-hsien+science19,jung+25acsnano} To confirm this hypothesis, we investigate in detail the band alignment of the \ce{CN4PMI}-\ce{WSe2} hybrid interfaces from DFT. 

\subsection{Computational Analysis}
\begin{figure}
    \centering
\includegraphics[width=0.5\textwidth]{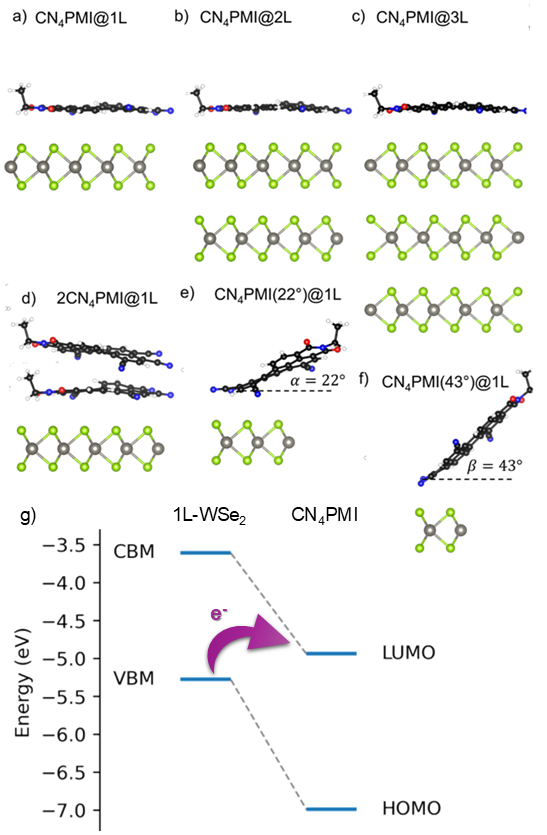}
    \caption{Ball-and-stick representations, created with the visualization software VESTA,\cite{Momma-db5098} of the unit cells of the investigated hybrid interfaces, including a single \ce{CN4PMI} adsorbed flat on a) monolayer, b) bilayer, and c) trilayer \ce{WSe2}, d) \ce{CN4PMI} molecules adsorbed on \ce{WSe2} monolayer and a single \ce{CN4PMI} molecule adsorbed on monolayer \ce{WSe2} with an angle e) $\alpha=22^{\circ}$ and f) $\beta = 43^{\circ}$.
    C, N, O, H, W, and Se atoms are depicted in black, blue, red, white, gray, and green, respectively. g) Schematic representation of the type-II level alignment between monolayer \ce{WSe2} and the \ce{CN4PMI} molecule. The purple arrow indicates the direction of electron transfer.}
    \label{fig:systems}
\end{figure}  

We model the \ce{CN4PMI}-\ce{WSe2} heterostructure assuming six different setups with varying numbers of layers and molecular orientations. We consider three configurations with \ce{CN4PMI} adsorbed flat on a \ce{WSe2} monolayer (Fig.~\ref{fig:systems}a) or with an angle of 22$^{\circ}$ and 43$^{\circ}$ (Fig.~\ref{fig:systems}e and Fig.~\ref{fig:systems}f, respectively). To address the influence of the organic film thickness, we additionally simulate two \ce{CN4PMI} lying flat on monolayer \ce{WSe2} (Fig.~\ref{fig:systems}d), while to confirm the robustness of the charge-transfer mechanism with respect to the substrate, we investigate \ce{CN4PMI} adsorbed flat on bilayer and trilayer \ce{WSe2}, see Fig.~\ref{fig:systems}b and c, respectively. 

The supercell dimensions reflect three distinct interaction regimes (Fig.~S5). Isolated molecules adsorbed flat on a $4 \times 5 \times 1$ \ce{WSe2} supercell minimize lateral intermolecular interactions and isolate the electronic coupling between dye and substrate. A dilute $\pi$-$\pi$ interaction regime allows evaluating the onset of collective molecular effects. Finally, a compact packing obtained on a $3 \times 2 \times 1$ \ce{WSe2} supercell simulates high-density molecular coverage. Intermolecular distances in these models are inherently anisotropic. This choice is physically motivated by the tendency of rylene derivatives to form one-dimensional $\pi$-$\pi$ stacked columns or J-aggregates upon physisorption.\cite{chen+15cr} Furthermore, in LB films, these dyes often adopt complex packing motifs, such as antiparallel head-to-tail arrangements,\cite{hupf+21langmuir} to minimize collective dipole-dipole repulsion. By varying the packing density, lattice anisotropy, and molecular orientation, we ensure a comprehensive sample characterization mirroring the structural complexity and varied molecular motifs of the considered \ce{CN4PMI-WSe2} heterostructures synthesized under ambient conditions.

\begin{table}
\centering
\caption{Adsorption energy, minimal distance ($d$) between adsorbed molecule and substrate, charge transfer (CT) from \ce{WSe2} to the organic layer, and energy gap of the hybrid interfaces modeled in this work}
\label{tab:Energy_abs-Gap}
\begin{tabular}{lcccc}
\hline \hline
System & E$_{ads}$ [eV] & $d$ [\AA{}] & CT [$e$] &E$_{gap}$ [eV] \\ \hline \hline
 \ce{CN4PMI}@1L &-2.42 & 3.52 & 0.23& 0.50     \\\hline
 \ce{CN4PMI}@2L &-2.46 & 3.59 & 0.22 & 0.47   \\ \hline
 \ce{CN4PMI}@3L &-2.47 & 3.48 & 0.22 &0.41   \\\hline
 2(\ce{CN4PMI})@1L &-2.60 & 3.65 & 0.31 & 0.00 \\\hline
 \ce{CN4PMI}(22$^{\circ}$)@1L &-1.15 & 3.16 & 0.07& 0.00\\\hline
 \ce{CN4PMI}(43$^{\circ}$)@1L &-0.58& 2.70 & 0.07 &0.00 \\\hline \hline
                            
\end{tabular}
\end{table}

To evaluate the interactions between adsorbed molecules and the underlying \ce{WSe2} substrate, we calculate from DFT the adsorption energy as the difference between the total energy of the hybrid interface and the total energies of its separate constituents in their optimized geometries: 
\begin{equation}\label{eq_ads}
    E_{\text{ads}} = E_{\text{hybrid}} - nE_{\text{\ce{WSe2}}} - mE_{\text{dye}}.
\end{equation}
In Eq.~\eqref{eq_ads}, $n$ and $m$ indicate the number of TMD and organic layers, respectively, that are present in each considered heterostructure. As summarized in Table~\ref{tab:Energy_abs-Gap}, all computed adsorption energies are negative and have a magnitude larger than 0.5~eV, suggesting favorable adsorption in all configurations. A careful inspection of these results reveals important differences according to the interface morphology. When the molecules lie flat on \ce{WSe2}, the adsorption energies assume the largest values, of the order of $-2.50$~eV, due to the strong interactions between the extended carbon-conjugated network of \ce{CN4PMI} and the TMD surface. This mechanism is further testified by the interfacial distance $d \approx 3.5$~\AA{}, which is almost insensitive to the number of TMD layers (Table~\ref{tab:Energy_abs-Gap}). 

The adsorption of two molecules on a \ce{WSe2} monolayer is more favorable by about 100~meV than single-molecule adsorption, despite the strong intermolecular couplings incrementing the distance between molecule and substrate to 3.65~\AA{} (Table~\ref{tab:Energy_abs-Gap}). Non-recumbent molecules adsorb less favorably on \ce{WSe2}, due to the substantial reduction of $\pi$-$\pi$ interactions between adsorbate and substrate. In these configurations, where the \ce{CN4PMI} forms an angle of 22$^{\circ}$ and 43$^{\circ}$ with \ce{WSe2}, the interaction with the TMD occurs mostly through the N atom at the edge, leading to a reduced separation as the angle increases. The adsorption energy decreases accordingly, down to -1.15~eV and -0.58~eV with angles of 22$^{\circ}$ and 43$^{\circ}$, respectively (Table~\ref{tab:Energy_abs-Gap}). 

Additional information about substrate-molecule interactions in these heterostructures can be gained by evaluating the degree of interfacial charge transfer (CT) from partial charge analysis, performed here adopting the Hirshfeld scheme.\cite{hirs77tca} As summarized in Table~\ref{tab:Energy_abs-Gap}, the CT trends follow those of the adsorption energies and minimal distances $d$. When the dyes are adsorbed flat on \ce{WSe2}, they withdraw 0.22~$e$, regardless of the number of underlying TMD layers. Two molecules adsorbed on \ce{WSe2} monolayer attract more than 0.3~$e$, confirming the strongest interaction obtained in this setting. Intermolecular CT in this heterostructure amounts to 0.05~$e$, which, interestingly, almost entirely compensates the difference with the CT induced by a single adsorbed dye. This result suggests that intermolecular interactions enhance interfacial couplings. Conversely, non-recumbent molecular adsorption leads to an electron transfer of only 0.07~$e$, reiterating the key role of $\pi$-interactions, which are minimized there.

\begin{figure*}
    \centering
\includegraphics[width=.95\textwidth]{newSOC-PDOS-rev.png}
    \caption{Projected densities of state (PDOS) of \ce{CN4PMI} adsorbed on \ce{WSe2} in all considered configurations: a) \ce{CN4PMI}:1L, b) \ce{CN4PMI}:2L, c) \ce{CN4PMI}:3L, d) 2(\ce{CN4PMI}):1L, e) \ce{CN4PMI}(22$^{\circ}$):1L, and f) \ce{CN4PMI}(43$^{\circ}$):1L. The contributions from molecule (purple) and \ce{WSe2} (orange) in the heterostructure are indicated by solid curves, while the results obtained for the isolated constituents are plotted by dashed curves (\ce{CN4PMI} in magenta and \ce{WSe2} in light orange). A 50~meV Lorentzian broadening is applied to all curves for visualization which causes an apparent overlap to the Fermi energy (E$_F$, marked by a dotted bar) in panel b). The energy scale is referenced to the vacuum level. }
    \label{fig:PDOS}
\end{figure*}

The non-negligible CT between \ce{CN4PMI} and \ce{WSe2}, which becomes larger with increasing interfacial interactions between the building blocks, is further confirmed by the analysis of the  PDOS (Fig.~\ref{fig:PDOS}). All heterostructures are thus characterized by a type-II level alignment, with the highest occupied level corresponding to the valence-band maximum (VBM) of \ce{WSe2} and the lowest unoccupied state by the LUMO of \ce{CN4PMI}. The LUMO of the molecule is found within the \ce{WSe2} gap in all considered configurations, a clear signature of $p$-doping induced by the molecule (Fig.~\ref{fig:systems}g). 

The size of the fundamental gap varies with the interface morphology. As reported in Table~\ref{tab:Energy_abs-Gap}, a single dye adsorbed flat on monolayer \ce{WSe2} is characterized by the largest gap of 0.5~eV. This magnitude decreases with the number of underlying TMD layers, leading to 0.47~eV (0.41~eV) for the bilayer (trilayer) substrate. Detailed inspection of the PDOS reveals that the largest band gap in the \ce{CN4PMI}@1L heterostructure is due to the maximized upshift of the LUMO energy compared to its value in the isolated molecule (Fig.~\ref{fig:PDOS}a). The magnitude of this shift decreases with the number of underlying \ce{WSe2} layers (Fig.~\ref{fig:PDOS}b,c) as a result of reduced polarization induced by the substrate. It is worth noting that the band edges of \ce{WSe2} monolayer undergo almost no shift in the presence of a single adsorbate (Fig.~\ref{fig:PDOS}a). 

In the hybrid interfaces with bilayer and trilayer substrates, both the VBM and the conduction band minimum (CBm) of \ce{WSe2} upshift by $\sim$100~meV upon \ce{CN4PMI} adsorption. The highest-occupied molecular orbital (HOMO) of the dye appears around -7~eV in the PDOS reported in Fig.~\ref{fig:PDOS}a-c. Similar to the LUMO, its energy undergoes an increasing upshift with decreasing number of \ce{WSe2} substrate layers. However, neither the HOMO nor the LUMO of \ce{CN4PMI} carries signatures of hybridization, in line with the rationale elaborated for non-functionalized rylene dyes on TMDs.\cite{krum+21es,mela+22pccp} On the other hand, as expected,\cite{krum+21es} the higher unoccupied molecular orbitals appear hybridized with the substrate, as indicated by the modified shape of the corresponding band appearing between -3.5~eV and -2.5~eV in the isolated and adsorbed dye (Fig.~\ref{fig:PDOS}a-c, dashed magenta \textit{vs.} solid purple curves).

The stronger interactions induced by two $\pi$-stacked dyes~\cite{elma+24jpca} adsorbed on monolayer \ce{WSe2} are reflected in the electronic structure of the corresponding interface (Fig.~\ref{fig:PDOS}d). As a result of strong doping, which sizeably upshifts the VBM of \ce{WSe2} by about 300~meV, this system has a vanishing gap (Table~\ref{tab:Energy_abs-Gap}). The strong interfacial interactions also lead to a splitting of both the HOMO and LUMO of the dyes, which become energetically close to those in an isolated dimer (Fig.~S4). Hybridization and polarization effects are quite pronounced in the unoccupied region of both constituents. Similar to the valence band, the conduction band of the TMD is upshifted by more than 300~meV, while the unoccupied molecular orbitals are substantially reorganized upon adsorption (Fig.~\ref{fig:PDOS}d).

Non-recumbent adsorption, while being energetically less favorable and leading to reduced CT than the flat configurations, induces non-negligible modifications in the electronic structure of both constituents (Fig.~\ref{fig:PDOS}e,f).
While the LUMO of \ce{CN4PMI} is almost unaffected by the presence of the substrate, undergoing an upshift of a few tens of meV only, the HOMO strongly hybridizes with the valence bands of \ce{WSe2} (Fig.~\ref{fig:PDOS}e). The electronic structure of the TMD is subject to a sizeable upshift in both the conduction and valence region, which closes the gap of the heterostructure (Table~\ref{tab:Energy_abs-Gap}). When the angle between \ce{CN4PMI} and \ce{WSe2} increases to 43$^{\circ}$, the changes in the electronic structure become even larger (Fig.~\ref{fig:PDOS}f). The LUMO of the molecule is smeared and energetically downshifted, overlapping with the upshifted VBM of \ce{WSe2}. The HOMO of \ce{CN4PMI} is also smeared upon adsorption, while the higher unoccupied levels are decreased in energy by $\sim$250~meV. A similar shift but toward higher energies is present in both the valence and conduction band of \ce{WSe2} (Fig.~\ref{fig:PDOS}f).

This transition highlights how the electrostatic environment, modulated by the molecular orientation and packing density, dictates the electronic properties of the system. In these more compact tilted configurations, the collective dipole moment of the functionalized rylene dyes, particularly the highly polar cyano-groups, significantly modifies the local electrostatic potential. This induces a shift in the molecular energy levels relative to the TMD bands, leading to the observed closure of the interfacial energy gap. Since the CT in the non-recumbent configurations is considerably smaller than upon flat molecular adsorption (Table~\ref{tab:Energy_abs-Gap}), we conclude that these strong electronic modifications at the interface are due to polarization and local interactions with the N atom at the edge of \ce{CN4PMI}, mostly interacting with the upper Se atomic layer in the TMD.

\section{Discussion and Conclusions}
Quenched PL observed in our experiments is directly related to the strong electronic interactions between \ce{CN4PMI} dyes and \ce{WSe2}, as revealed by DFT calculations. As extensively discussed in the literature on a variety of interfaces with organic molecules~\cite{khan+20ass,amst+21jacs,herr15jpcc} and inorganic substrates,\cite{muns+24jpcl,zhu+25jpcc} CT and doping are considered detrimental to the intrinsic optical activity of TMDs. Both effects manifest themselves in a type-II level alignment, leading to a substantial band-gap reduction in the heterostructures compared to the pristine semiconductors. Our comprehensive first-principles analysis confirms this hypothesis across various interfacial morphologies. Although flat molecular adsorption is energetically and electronically superior, even non-recumbent orientations induce modifications leading to sizable PL quenching.
These results underscore the role of molecular design in modulating the interfacial electronic landscape.\cite{gera+25am} In particular, the use of electron-deficient \ce{CN4PMI}, featuring multiple cyano substituents, enables efficient charge transfer and energetic alignment with \ce{WSe2}. The planarity and high electron affinity of the dye are key features that enable efficient charge transfer and fine-tuning of the TMD band structure via non-covalent interactions.

In conclusion, we attribute the observed PL quenching in \ce{WSe2} decorated by functionalized rylene dyes to $p$-doping induced by the molecules. \textit{Ab initio} calculations, performed for varying molecular coverage and orientation, confirm the robustness of this predicted mechanism. The adsorbed dyes consistently accept electrons from \ce{WSe2}, suggesting that in real samples, the available charge fills Se vacancy states, thereby suppressing PL. Beyond explaining this quenching within the broader context of TMD heterostructures, our findings demonstrate a clear pathway for tailoring the electronic structure of these low-dimensional semiconductors via controlled organic functionalization, opening avenues for advanced optoelectronics, nonlinear optics, and quantum technologies. 
In particular, the robust $p$-doping induced by the dye provides a viable pathway for the fabrication of $p$-type field-effect transistors, addressing the long-standing challenge of controllable $p$-type functionalization in TMDs. Finally, the extreme sensitivity of \ce{WSe2} to molecular decoration highlights the potential of these hybrid interfaces as active components for high-contrast chemical sensing.

\section*{Author contributions}
\textbf{A.M.V.}: Investigation (equal), Data Curation (lead), Visualization (equal), Writing -- Original Draft (lead); \textbf{T.K.} Investigation (equal), Writing -- Review \& Editing (supporting); \textbf{M.T.}: Investigation (supporting), Validation (lead), Data Curation (supporting), Visualization (equal), Writing -- Original Draft (supporting); \textbf{S.J.F.} Investigation (supporting), Validation (supporting), Writing -- Review \& Editing (supporting); \textbf{O.U.} Investigation (supporting); \textbf{K.P.} Conceptualization (equal), Resources (equal); Supervision (equal), Funding acquisition (equal), Writing -- Review \& Editing (supporting); \textbf{M.P.} Resources (equal); Supervision (equal), Funding acquisition (equal), Writing -- Review \& Editing (supporting); \textbf{G.S.} Conceptualization (equal), Resources (equal); Supervision (equal), Funding acquisition (equal), Writing -- Review \& Editing (supporting); \textbf{C.C.} Conceptualization (equal), Resources (equal); Supervision (equal), Funding acquisition (equal), Writing -- Review \& Editing (lead).

\section*{Conflicts of interest}
The authors declare no competing financial or personal conflict of interest regarding this work, except that the company SciClus GmbH \& Co. KG (founded by M.P.) provided IT infrastructure free of charge for the research. However, this company derived no financial or non-financial benefit from the research and was not involved in the study's design, data collection, analysis, or interpretation, or in the writing of the manuscript.

\section*{Data availability}
All data produced in this work are available free of charge at
\url{https://doi.org/10.5281/zenodo.17294219}.

\section*{Acknowledgements}
The authors thank Benedikt Callies for support with the cyclic voltammetry measurements. 
This work was funded by the German Federal Ministry of Education and Research (Professorinnenprogramm III), the State of Lower Saxony (Professorinnen f\"ur Niedersachsen), and  the German Research Foundation (DFG) via CRC 1375 NOA (project number 398816777, subprojects A8 and B5) and FOR5301 ``FuncHeal'' (project number 455748945, subprojects P3 and P4). Computational resources were provided by the North-German Supercomputing Alliance (HLRN), project bep00132, and by the high-performance computing cluster CARL at the University of Oldenburg, funded by the German Research Foundation (Project No. INST 184/157-1 FUGG) and by the Ministry of Science and Culture of Lower Saxony.  



\balance



\providecommand*{\mcitethebibliography}{\thebibliography}
\csname @ifundefined\endcsname{endmcitethebibliography}
{\let\endmcitethebibliography\endthebibliography}{}

\bibliographystyle{rsc} 
\end{document}